\documentclass[11pt]{article}  
\def\baselinestretch{1.1}
\usepackage{amsmath}
\usepackage{amsthm}
\usepackage{amsfonts}
\usepackage{amssymb}
\usepackage{times}
\newcommand{\beq}[1]{ 
 \begin{equation}\label{#1}}
\def\PP{{\mathbb P}}
\def\ot{\otimes}
\def\om{\omega}
\def\mus{{\mu^{\vphantom{2}}_S}}
\def\pis{\pi_{\scriptscriptstyle S}}
\newcommand{\fr}[2]{{\textstyle \frac{#1}{#2} }}
\def\klein{\scriptscriptstyle}

\newcommand{\0}{{\mathfrak 0}}
\newcommand{\1}{{\mathfrak 1}}
\newcommand{\2}{{\mathfrak 2}}
\newcommand{\3}{{\mathfrak 3}}

\def\p1{{\scriptscriptstyle +1}}

\newcommand{\rf}[1]{(\ref{#1})}
\newcommand{\SA}{{\mathsf A}}
\newcommand{\SB}{{\mathsf B}}
\newcommand{\SC}{{\mathsf C}}

\newcommand{\la}{\lambda}

\renewcommand{\=}[1]{\stackrel{(\ref{#1})}{=}}

\newcommand{\SR}{{\mathsf R}}

\newcommand{\SK}{{\mathsf K}}
\newcommand{\SH}{{\mathsf H}}
\newcommand{\SP}{{\mathsf P}}

\newtheorem{propn}{\bfseries Proposition}
\newtheorem{lem}{\bfseries Lemma}
\newtheorem{conj}{\bfseries Conjecture}

\theoremstyle{remark}
\newtheorem{rem}{Remark}
\newcommand{\scp}[2]{\langle{#1},#2\rangle}

\begin{document}
\thispagestyle{empty}
{\small
November 2009  \hfill
 \strut\hfill  DESY 09--206 \\
 }
\vspace*{-1mm}
\begin{center}
{\Large\bf
Non--Hermitian spin chains with inhomogeneous coupling
} \\ [3mm]
{\sc Andrei G. Bytsko}    \\ [2mm]
{ \small
   Steklov Mathematics Institute,
 Fontanka 27, 191023, St.~Petersburg, Russia \\ [1mm]
 DESY Theory Group, Notkestrasse 85, D-22603, Hamburg, Germany
} \\ [2.5mm]
{ }

\end{center}
\vspace{1mm}
\begin{abstract}
\noindent
An open $U_q(sl_2)$--invariant spin chain of spin $S$ and
length $N$ with inhomogeneous coupling is investigated
as an example of a non--Hermitian (quasi--Hermitian) model.
For several particular cases of such a chain, the ranges of
the deformation parameter
$\gamma$ are determined for which the spectrum of the model
is real. For a certain range of $\gamma$,
a universal metric operator is constructed and thus
the quasi--Hermiticity of the model is established.
The constructed metric operator is non--dynamical, its
structure is determined only by the symmetry of the model.
The results apply, in particular, to all known homogeneous
$U_q(sl_2)$--invariant integrable spin chains with
nearest--neighbour interaction.
In addition, the most general form of a metric operator for
a quasi--Hermitian operator in finite dimensional space is
discussed.

\end{abstract}
%


%
\section*{Introduction}

A bounded linear operator $\SH$ in a complex Hilbert space
$\mathfrak H$ equipped with the inner product $\scp{x}{y}$ is
said to be {\em symmetrizable} if there exists
a Hermitian operator $\eta$ such that $\eta \,{\neq}\, 0$ and
\beq{Heta}
   \eta\, \SH = \SH^*  \eta  \,.
\end{equation}
Symmetrizable operators have been studied in
mathematical literature since long ago \cite{Za,Re,He,Di,S1,S2}.
Following Dieudonne \cite{Di}, we will say that a symmetrizable
operator $\SH$ is {\em quasi--Hermitian} if the symmetrizing
operator $\eta$ is positive definite.

If $\eta$ is invertible then a quasi--Hermitian operator $\SH$
is similar to a Hermitian one and hence it has a real spectrum
(the spectrum of $\SH$ can be not entirely real if $\eta$ is positive
definite but not invertible, see~\cite{Di,S2}). This enables
an interpretation~\cite{SGH} of an irreducible set of
quasi--Hermitian operators as quantum mechanical observables
if they share a common symmetrizing operator~$\eta$.
In this context $\eta$
is called a {\em metric operator} since the observables become
Hermitian operators with respect to the modified inner product
$\scp{x}{y}_\eta\equiv\scp{x}{\eta y}$.
Interesting motivating examples of non--Hermitian operators
with a real spectrum are the Hamiltonian of the lattice
Reggeon field theory \cite{CS}, the Hamiltonian of the Ising
quantum spin chain in an imaginary magnetic field~\cite{Ge},
the Hamiltonians of affine Toda field theories with an imaginary
coupling constant \cite{Ho},
and the Schr\"odinger operator with an imaginary cubic
potential~\cite{BZ}. The latter example was generalized
\cite{BB2} to a large class of symmetrizable Hamiltonians
possessing the PT (parity and time--reversal) symmetry and
having, according to Wiegner's theory \cite{Wi} of anti--unitary
operators, (partially) real spectra.
Since then a lot of research in physical literature
has been devoted to symmetrizable and, in particular, quasi--Hermitian
Hamiltonians, leading to the construction of numerous interesting
examples and the (re)discovery of many mathematical aspects; see
\cite{Be,M2} for reviews.

The Hamiltonian $\SH$ of a physical model is often given by the
sum or, more generally, a linear combination of local Hamiltonians
$\SH_n$, $n\,{=}\,1,{\ldots},N$ with real coefficients
(coupling constants)
\beq{HN}
    \SH = \sum_{n=1}^N  a_n \, \SH_n  \,,
    \qquad a_n \in \mathbb{R} \,.
\end{equation}
Here we face an immediate difficulty not present in the theory
of Hermitian operators: no general criterion is known that would
determine whether $\SH$ is a quasi--Hermitian operator given that
all $\SH_n$ are quasi--Hermitian operators
(it is not assumed that they share a common symmetrizing operator).
This problem naturally arises for Hamiltonians of
various spin chains where the interaction between
adjacent sites is described by quasi--Hermitian operators.
For instance, the reality of spectra and the existence of metric
operators for such compound chains have been investigated for
the Ising chain in an imaginary magnetic field \cite{Ge,CF},
the Jordanian twist of the Heisenberg chain \cite{KS},
and the homogeneous XXZ model of spin $\frac 12$~\cite{KW}.
In the present paper we will address the problem of quasi--Hermiticity
for an open spin chain of spin $S$ with nearest--neighbour
Hamiltonians $\SH_n$ having most general form
respecting $U_q(sl_2)$ symmetry.

The paper is organized as follows. In Section~1.1, we provide the
necessary facts about quasi--Hermitian operators, and in Section~1.2,
discuss the most general form of a metric operator. In Section~2.1,
we recall the basic notions related to the quantum algebra $U_q(sl_2)$,
discuss the phenomenon of non--Hermiticity for the tensor product
of its representations in the case of $q \,{=}\,e^{i\gamma}$,
$\gamma \,{\in}\,\mathbb R$, and
introduce an open $U_q(sl_2)$--invariant spin chain of length $N$
with inhomogeneous coupling. In Sections~2.2 and~2.3, we investigate
the reality of spectra of particular cases of such a chain
for $N \,{=}\,3,4,5$ by considering the minimal polynomials of
the corresponding Hamiltonians. Extrapolating our results,
we formulate two conjectures on the range of
$\gamma$ in which the spectrum is real. In Section~2.4, we construct
a multi--parametric family of universal, i.e. independent of coupling
constants, symmetrizing operators for the most general
$U_q(sl_2)$--invariant open spin chain
with a nearest--neighbour interaction. The construction exploits
solely the quantum algebraic symmetry of the model and is
formulated in terms of related algebraic objects such as
the R--matrix and the comultiplication. For a one--parametric
subfamily of symmetrizing operators, we determine
the range of $\gamma$ in which it contains positive definite
operators and thus the Hamiltonian of the model
is quasi--Hermitian. In Conclusion we summarize and briefly
discuss our results. Appendix contains
proofs of the statements given in the main text and some technical
details on R--matrices and projectors on irreducible subspaces
in tensor products.

\section{Quasi--Hermitian operators and metric operators}
\subsection{Preliminaries}
\label{PRE}

Consider the eigenvalue problem for a quasi--Hermitian operator $\SH$,
\beq{Heig}
  \SH \, \omega_j = \la_j \, \om_j \,,
    \qquad \scp{\om_j}{\om_j}=1 \,.
\end{equation}
Let $\{\om_j\}$  be the set of normalized
eigenvectors of $\SH$ and $\mathrm{ Spec}(\SH) \equiv \{\la_j\}$
be the set of the corresponding eigenvalues.
Here and below we will restrict our consideration to the case
of finite dimensional Hilbert space,
$d \equiv \mathrm{dim}\, \mathfrak H \,{<}\,\infty$.
In this case, the metric operator $\eta$ is invertible
and the quasi--Hermitian operator $\SH$ is similar to
a Hermitian operator
$\eta^{\klein \frac 12} \SH \eta^{-\klein \frac 12}$.
Whence it is immediate that
$\mathrm{Spec}(\SH) \subset \mathbb{R}$,
and the set $\{\om_j\}$ is a complete set of vectors in~$\mathfrak H$.

\rem\label{eta0}
The converse is also true, see~\cite[Thm.\,3.3]{S1}:
if a linear operator $\SH$ in a finite
dimensional complex Hilbert space $\mathfrak H$ has a real spectrum
and the set $\{\om_j\}$ of its eigenvectors is complete, then $\SH$
is quasi--Hermitian. A metric operator for a given $\SH$ can be
constructed as follows (see e.g. \cite{M1}): take an arbitrary
orthonormal basis $\{e_j\}$ in~$\mathfrak H$ and define
a linear operator $\Omega$ such that $\Omega \omega_j \,{=}\, e_j$.
Then $\Omega$ is invertible and
$\SH_0 \,{=}\, \Omega \SH \Omega^{-1}$ is Hermitian.
Whence it follows that $\eta_{\klein 0} \,{=}\, \Omega^* \Omega$
is a metric operator for~$\SH$. Note that $\eta_{\klein 0}$ does
not actually depend on the choice of the basis~$\{e_j\}$.

\rem
In physical literature on PT--symmetric models \cite{BBJ,Be,M2,AF},
one considers also {\em pseudo--Hermitian} operators,
i.e. symmetrizable operators for which $\eta$ is invertible
but not positive definite.
Pseudo--Hermiticity of $\SH$ implies only that, if
$\la\,{\in}\,\mathrm{Spec}(\SH)$, then
$\bar{\la}\,{\in}\,\mathrm{Spec}(\SH)$,
as for instance in the case of
$\SH=\left(\begin{smallmatrix}i&0\\0&-i\end{smallmatrix}\right)$,
$\eta=\left(\begin{smallmatrix}0&1\\1&0\end{smallmatrix}\right)$.
Furthermore, the set of the eigenvectors of a pseudo--Hermitian operator
is not necessarily a complete set of vectors in~$\mathfrak H$,
as another simple example demonstrates:
$\SH=\left(\begin{smallmatrix}1&1\\0&1\end{smallmatrix}\right)$,
$\eta=\left(\begin{smallmatrix}0&1\\1&0\end{smallmatrix}\right)$.


The eigenvectors $\{\om_j\}$ of a quasi--Hermitian operator
$\SH$ provide a {\em non--orthogonal} basis in~$\mathfrak H$.
Consider the corresponding {\em Gram matrix} $G$
with entries
$G_{kn} = \scp{\om_k}{\om_n}$. 
The matrix $G$ is invertible, Hermitian (with respect to the
conjugate transpose operation), and positive definite. 
The set of vectors $\{\tilde{\om}_j\}$, where
$\tilde{\om}_j \,{=}\, \sum_{n =1}^d (G^{-1})_{nj} \om_n$, provides
another non--orthogonal basis in~$\mathfrak H$.
Its Gram matrix is $G^{-1}$.
The bases $\{\om_j\}$ and $\{\tilde{\om}_j\}$
form a {\em bi--orthogonal} system:
\beq{bio}
 \scp{\om_k}{\om_j} = \delta_{kj} \,, \qquad
 \scp{\om_k}{\tilde{\om}_j} = \delta_{kj} \,, \qquad
 \scp{\tilde{\om}_k}{\tilde{\om}_j}= (G^{-1})_{kj} \,.
\end{equation}

\begin{rem}\label{omnorm}
Note that $\tilde{\om}_j$ are, in general, {\em not normalized}.
Indeed, positive definiteness of $G^{-1}$ implies only that
$ (G^{-1})_{jj} \,{>}\, 0$ for all~$j$.
\end{rem}

Any vector $x\in\mathfrak H$ defines a linear functional
$x^\dagger : \mathfrak H \mapsto \mathbb C$ such that
$x^\dagger (y)=\scp{x}{y}$. Since $\{\om_j\}$ and $\{\tilde{\om}_j\}$
are bases in~$\mathfrak H$, any linear operator $\SA$ acting
in $\mathfrak H$ can be written in the form
\beq{Asym}
\SA=\sum_{k,n=1}^d O(\SA)_{kn} \om_k \, \om^\dagger_n =
 \sum_{k,n=1}^d \tilde{O}(\SA)_{kn}
    \tilde{\om}_k \, \tilde{\om}^\dagger_n \,,
\end{equation}
where $O(\SA)$ and $\tilde{O}(\SA)$ are complex matrices
(we will call them {\em symbols} of~$\SA$). It is useful to
observe that $O(\SA^*)=\bigl(O(\SA)\bigr)^*$,
$\tilde{O}(\SA^*)=\bigl(\tilde{O}(\SA)\bigr)^*$, and
\begin{align}
\label{OOt1}
 O(\SA\,\SB)=O(\SA)\,G\,O(\SB) \,,&  &&
\tilde{O}(\SA\,\SB)=\tilde{O}(\SA)\,G^{-1}\tilde{O}(\SB) \,, \\
\label{OOt2}
  \tilde{O}(\SA) = G O(\SA) G \,,& &&
\tilde{O}(\SA) \, O(\SA^{-1}) = E \,,
\end{align}
where $E$ is the identity matrix, and the last relation
makes sense if $\SA$ is invertible.

Let $\SP_j$ and $\tilde{\SP}_j$ denote projectors in $\mathfrak H$
on $\om_j$ and $\tilde{\om}_j$, respectively, i.e.
$\SP_j \, \om_k = \delta_{jk} \, \om_j$ and
$\tilde{\SP}_j \, \tilde{\om}_k = \delta_{jk} \, \tilde{\om}_j$.
Relations \rf{bio} imply that these projectors are given by
\beq{PPj}
  \SP_j = \om_j \, \tilde{\om}_j^\dagger
 = \sum_{n= 1}^d  (G^{-1})_{jn} \,
    \om_j \, \om_n^\dagger =
 \sum_{n= 1}^d   G_{nj} \,
    \tilde{\om}_n \, \tilde{\om}_j^\dagger \,, \qquad
 \tilde{\SP}_j = \SP^*_j = \tilde{\om}_j \, \om_j^\dagger \,.
\end{equation}
The resolutions of the unity,
$\sum\limits_{j = 1}^d \SP_j \,{=}\, {\sf 1} \,{=}
\sum\limits_{j = 1}^d \SP^*_j$, are due to the completeness
of the sets $\{\om_j\}$ and~$\{\tilde{\om}_j\}$.

\subsection{General form of metric operator}

Consider a quasi--Hermitian operator $\SH$ which has $d'\leq d$
distinct eigenvalues $\{\la_j\}$ with multiplicities $\mu_j \geq 1$,
so that we have $\sum_{j=1}^{d'} \mu_j =d$.
The eigenvectors corresponding to a given eigenvalue $\la_j$
span the subspace $\mathfrak H_j \subset \mathfrak H$.
Let $\{\om_{j,k}\}$, $k \,{=}\,1,{\ldots},\mu_j$ be a basis
of $\mathfrak H_j$ (it is not unique if $\mu_j \,{>}\, 1$)
and let $\SP_{j,k}$ denote the projector on~$\om_{j,k}$.

\begin{propn}\label{etadeg}
a)
For a quasi--Hermitian operator $\SH$ which has the spectrum
$\{\la_j\}$
with multiplicities~$\mu_j$, fix some basis $\{\om_{j,k}\}$ in each
subspace $\mathfrak{H}_j$. Then, for this $\SH$, the most general form
of a metric operator and its inverse is the following
\beq{etappdeg}
\eta=\sum_{j= 1}^{d'} \sum_{k,n= 1}^{\mu_j}
    \bigl(\Phi_j\bigr)_{kn} \, \tilde{\om}_{j,k} \,
    \tilde{\om}_{j,n}^\dagger \,, \qquad
\eta^{-1} =\sum_{j= 1}^{d'} \sum_{k,n= 1}^{\mu_j}
 \bigl(\Phi^{-1}_j\bigr)_{kn} \, \om_{j,k} \, \om_{j,n}^\dagger \,,
\end{equation}
where $\Phi_j$ are arbitrary Hermitian positive definite
matrices of size $\mu_j{\times}\mu_j$.


\noindent
b)
For a quasi--Hermitian operator $\SH$ which has the spectrum
$\{\la_j\}$
with multiplicities~$\mu_j$, take some metric operator~$\eta$.
Then there exists a choice of bases $\{\om_{j,k}\}$
of subspaces $\mathfrak{H}_j$ such that the given operator
$\eta$ and its inverse are given by
\beq{etapp}
\eta
 = \sum_{j=1}^{d'} \sum_{k= 1}^{\mu_j}
    \Phi_{j,k} \, \SP^*_{j,k} \, \SP_{j,k} \,, \qquad
 \eta^{-1}
 = \sum_{j=1}^{d'} \sum_{k= 1}^{\mu_j}
    \widetilde{\Phi}_{j,k} \,\SP_{j,k} \, \SP^*_{j,k} \,,
\end{equation}
where $\Phi_{j,k}$ are arbitrary positive numbers and
$\widetilde{\Phi}_{j,k}=
\bigl ((G^{-1})_{\{j,k\},\{j,k\}} \, \Phi_{j,k} \bigr)^{-1}$.
\end{propn}

\begin{rem}
It is natural to regard metric operators differing only by
a positive constant scalar factor as equivalent.
Thus, formulae \rf{etapp} describe $(d \,{-}\,1)$--parametric
families of operators.
If the spectrum of a quasi--Hermitian operator $\SH$ is simple,
then these formulae give the most general form of the
corresponding metric operator and its inverse.
\end{rem}

\begin{rem}\label{G0}
As noted in the previous Remark, the parts {\em a)} and {\em b)}
of Proposition~\ref{etadeg} are just different forms of the same
statement if the spectrum of $\SH$ is simple.
The difference appears if the spectrum
of $\SH$ is degenerate. Indeed, although any given metric operator
can be brought to the form \rf{etapp} which involves only the
projectors on the eigenvectors of $\SH$, this requires
a change of the basis in the Hilbert space {\em after} we have
chosen the metric operator. But if we work with a {\em fixed} basis,
then the most general form of a metric operator \rf{etappdeg} cannot
in general be re--expressed only in terms of the projectors
on the eigenvectors of $\SH$ if it
has a degenerate spectrum. This is so because
$\SP^*_{j,k}\SP_{j,n}=
 G_{\{j,k\},\{j,n\}} \tilde{\om}_{j,k} \tilde{\om}^\dagger_{j,n}$,
and the corresponding entry of the Gram matrix can be zero.
(In fact, it is zero, if we choose an orthonormal basis in the
subspace~${\mathfrak H}_j$.)
\end{rem}



\begin{rem}
If all $\Phi_j$ are identity matrices, then \rf{etappdeg}
yields the operator $\eta_{\klein 0}$ considered in Remark~\ref{eta0}.
Indeed, it easy to see that
$\Omega^{-1} = \sum_{j=1}^d \om_j \, e^\dagger_j$,
whence $\eta_{\klein 0}^{-1} = \Omega^{-1} (\Omega^*)^{-1} =
 \sum_{j=1}^d \om_j \, \om^\dagger_j$.
\end{rem}

\begin{rem}
If $\SH$ has a simple spectrum, we can rewrite formulae
\rf{etapp} using Eqs.~\rf{pfunh} into a form that does not use
eigenvectors explicitly:
\begin{align}
\label{etaHH}
\eta &= \sum_{j = 1}^d \Theta_{j} \,
 \Bigl( \prod_{n \neq j}^d (\SH^* - \la_n\,{\sf 1}) \Bigr)\,
 \Bigl( \prod_{m \neq j}^d (\SH - \la_m\,{\sf 1}) \Bigr) \,,\\
\label{etainvHH}
\eta^{-1} &= \sum_{j = 1}^d \widetilde{\Theta}_{j} \,
 \Bigl( \prod_{m \neq j}^d (\SH - \la_m\,{\sf 1}) \Bigr) \,
 \Bigl( \prod_{n \neq j}^d (\SH^* - \la_n\,{\sf 1}) \Bigr)\,,
\end{align}
where $\Theta_j$ are arbitrary positive numbers and
$\widetilde{\Theta}_j= \bigl( (G^{-1})_{jj} \, \Theta_j \bigr)^{-1}$.
\end{rem}

As an example, consider the following operator acting
in ${\mathbb C}^2$ (it is related to the Hamiltonian~(93)
in \cite{Be} by a change of variables which ensures
reality of the spectrum):
\beq{Hb}
   \SH=
 \begin{pmatrix}
        e^{i\theta} \, \sinh z & \sin\theta \, \cosh z \\
        \sin\theta \, \cosh z & e^{-i\theta} \, \sinh z\end{pmatrix} =
 (\sinh{z}) \, e^{i \theta \sigma_\3} +
  (\sin\theta \, \cosh z) \, \sigma_\1  \,, \qquad
  \theta, z \in \mathbb R .
\end{equation}
Here and below we use the standard notations for the Pauli matrices:
$\sigma_\1=\left(\begin{smallmatrix} 0 & 1 \\
  1 & 0 \end{smallmatrix}\right)$,
$\sigma_\2=\left(\begin{smallmatrix} 0 & -i \\
  i & 0 \end{smallmatrix}\right)$,
$\sigma_\3=\left(\begin{smallmatrix} 1 & 0 \\
  0 & -1 \end{smallmatrix}\right)$.
Operator \rf{Hb} is not Hermitian but has real eigenvalues
$\la_\pm=\cos\theta \, \sinh z \pm \sin\theta$.
Observe that its spectral resolution can be written in the
following form
\beq{Hbpp}
  \SH = \la_+ \SP_+ + \la_- \SP_- \,, \qquad
  \SP_{\pm} =  e^{-\frac{z}{2} \sigma_\2} \,
 \frac{ ( {\sf 1} \,{\pm}\, \sigma_\1 ) }{2} \,
 e^{\frac{z}{2} \sigma_\2} \,,
\end{equation}
which makes it obvious that $\SH=\Omega^{-1}\SH_0 \Omega$,
where $\Omega \,{=}\, e^{\frac{z}{2} \sigma_\2}$ and $\SH_0$
is Hermitian. Whence, by Remark~\ref{eta0}, we have
$\eta_{\klein 0} \,{=}\,\Omega^*\Omega \,{=}\,e^{z \sigma_\2}$,
whereas \rf{etapp} yields a one parametric family of metric
operators. Namely, taking
$\Phi_\pm = e^{\pm\varphi}/\!\cosh z$, where
$\varphi \in \mathbb R$, we obtain
\beq{etabb}
   \eta_\varphi =
   e^{\frac{z}{2} \sigma_\2 } \,
   e^{\varphi \sigma_\1 } \,
   e^{\frac{z}{2} \sigma_\2 }  \,.
\end{equation}
In this form, positive definiteness of $\eta_\varphi$ is
self--evident, and we recover $\eta_{\klein 0}$ for $\varphi \,{=}\,0$.

\section{Spin chains with inhomogeneous coupling}
\subsection{Spin chains with $U_q(sl_2)$ symmetry}
We will consider one dimensional lattice models (open chains
with free boundary conditions) which have $U_q(sl_2)$ symmetry.
Recall that the algebra $U_q(sl_2)$ has
the following defining relations
\beq{Uq}
 [E,F] = \fr{K^2- K^{-2}}{q-q^{-1}} , \qquad
 K E= q E K, \qquad KF = q^{-1} F K .
\end{equation}
A comultiplication consistent with these relations
can be chosen as follows:
\beq{DelEFK}
 \Delta(E) = E \ot K^{-1} + K \ot E \,, \quad
 \Delta(F) = F \ot K^{-1} + K \ot F \,,
 \quad \Delta(K) = K \ot K \,.
\end{equation}

Let $S$ be a positive integer or
semi--integer number, and let $q\,{=}\,e^{i\gamma}$, where
$\gamma \,{\in}\, \mathbb R$ and $2S|\gamma| \,{<}\, \pi$.
Let $V^S \,{\simeq}\, {\mathbb C}^{2S+1}$
be an irreducible highest weight $U_q(sl_2)$ module
and $\{\om_k\}_{k=-S}^S$ be its canonical orthonormal basis
in which $K$ is diagonalized.
We will consider the standard representation $\pis$ of
$U_q(sl_2)$ on $V^S$:
\beq{EFK}
\begin{aligned}
 \pis(E)\,\om_k &= \sqrt{[S{-}k][S{+}k{+}1]} \, \om_{k+1} , \\
 \pis(F)\,\om_k &= \sqrt{[S{+}k][S{-}k{+}1]} \, \om_{k-1} ,
\end{aligned}
\qquad
 \pis(K)\,\om_k = q^k \, \om_{k} ,
\end{equation}
where $[t] \equiv \frac{\sin \gamma t}{\sin \gamma}$.
In particular, $\pi_{\klein\frac{1}{2}}(E) \,{=}\,\sigma^+
 \,{\equiv}\, \frac{1}{2}(\sigma_\1 \,{+}\, i\sigma_2)$,
$\pi_{\klein \frac 12}(F) \,{=}\, \sigma^- \,{\equiv}\,
 \frac{1}{2}(\sigma_\1 \,{-}\, i \sigma_\2)$,
$\pi_{\klein \frac 12}(K) =e^{i \frac{\gamma}{2} \sigma_\3}$.
For $2S|\gamma| \,{<}\, \pi$, the non--zero matrix entries of
$\pis(E)$ and $\pis(F)$ are positive, and these matrices
are conjugate transposed to each other. Therefore,
Eqs.~\rf{EFK} can be regarded as a representation of the
algebra $U_q(sl_2)$ with the involution
\beq{EFK*}
 E^* =F \,, \qquad  F^* =E \,, \qquad K^* = K^{-1} \,.
\end{equation}
However, the algebra $U_q(sl_2)$ with such an involution is not
a Hopf $*$--algebra, i.e.,
$\bigl(\Delta(X)\bigr)^* \neq \Delta(X^*)$ in general.
Instead we have
$\bigl(\Delta(X)\bigr)^* = \PP \Delta(X^*) \PP$,
where $\PP$ is the operator of permutation of the tensor factors
in $U_q(sl_2)^{\otimes 2}$. This is the origin of non--Hermiticity
of models that will be considered below.

The comultiplication \rf{DelEFK} determines the decomposition
$V^S \,{\otimes}\, V^S \,{=}\, {\oplus}_{s=0}^{2S} V^s$,
where each $V^s$ is an irreducible $U_q(sl_2)$--submodule.
The inner product on $V^S$ gives rise to an inner product on
$V^S \,{\otimes}\, V^S$:
$\scp{\om_k \,{\ot}\, \om_{m}}{\om_{k'} \,{\ot}\, \om_{m'}}
=\delta_{kk'}\delta_{mm'}$.
A basis for $V^S \,{\otimes}\, V^S$ can be
taken to be $\{\om_{s,k}\}$, where $s\,{=}\,0,{\ldots},2S$,
and, for given $s$, vectors $\om_{s,k}$,
$k\,{=}\,-S,{\ldots},S$ comprise the canonical basis of~$V^s$.

An important difference between the cases $q \,{\in}\, \mathbb R$
and $|q| \,{=}\, 1$ is that in the latter case vectors from different
submodules
can be non--orthogonal. For instance, the
basis for $V^{\klein \frac 12} \,{\simeq}\, {\mathbb C}^2$
is $\om_{\klein \frac 12} \,{=}\,
 \bigl(\! \begin{smallmatrix} 1 \\ 0 \end{smallmatrix} \!\bigr)$,
$\om_{\klein -\frac 12} \,{=}\,
 \bigl(\! \begin{smallmatrix} 0 \\ 1 \end{smallmatrix} \!\bigr)$,
and the basis for
$V^{\klein \frac 12} \,{\ot}\,V^{\klein \frac 12} \,{=}\,
 V^0 \,{\oplus}\, V^1$ is
\begin{align}
\label{oms12}
 \om_{\klein 0,0}  = \fr{1}{\sqrt{\varkappa}} \!
 \left(\!\begin{smallmatrix} 0\\q^{ -\frac 12}
\\-q^{\frac 12}\\0\end{smallmatrix} \!\right)
 , \quad
 \om_{\klein 1,1} = \left(
 \begin{smallmatrix} 1\\0\\0\\0\end{smallmatrix} \right)
 , \qquad
 \om_{\klein 1,0} =  \fr{1}{\sqrt{\varkappa}} \!
 \left(\! \begin{smallmatrix} 0\\q^{\frac 12}\\
 q^{-\frac 12}\\0\end{smallmatrix} \! \right)
 , \qquad
 \om_{\klein 1,-1} =
 \left(\begin{smallmatrix} 0\\0\\0\\1\end{smallmatrix}\right) .
\end{align}
For $q \,{\in}\, {\mathbb R}$, these vectors are orthogonal, and
normalization requires to set $\varkappa \,{=}\, [2]$.
For $|q| \,{=}\, 1$, the vectors are normalized if
$\varkappa \,{=}\, 2$, and we have
$\scp{\om_{\klein 0,0}}{\om_{\klein 1,0}} \,{=}\, i \sin\gamma$.

\begin{rem}
Only those basis vectors from different submodules can be
non--orthogonal that have equal eigenvalues under the action of
$\SK_{12} \,{=}\, (\pis \,{\ot}\, \pis) \Delta(K)$.
Indeed, it follows from \rf{DelEFK} and \rf{EFK*} that
$\SK_{12}$ is unitary, $\SK^*_{12} \,{=}\, \SK^{-1}_{12}$.
Therefore, if $\SK_{12}\om \,{=}\,q^k \om$ and
$\SK_{12}\om' \,{=}\,q^{k'} \om'$, then
$\scp{\om'}{\SK_{12}\om} =
 q^k \scp{\om'}{\om}$ and hence
$q^{-k} \scp{\om}{\om'} =
 \scp{\om}{\SK^*_{12}\om'}=
 \scp{\om}{\SK^{-1}_{12}\om'}=q^{-k'} \scp{\om}{\om'}$,
which implies that
$q^k = q^{k'}$ if $\scp{\om}{\om'} \neq 0$.
\end{rem}

Let $\SP^{S,s}$ denote the projector onto the irreducible
submodule $V^s$ in $V^S \,{\otimes}\,V^S$.
Some details on the structure of these projectors are given in
Appendix~\ref{APP}.  In particular, the projectors $\SP^{S,s}$ are
not Hermitian but 
they are symmetrizable operators:
\beq{P*}
  \bigl( \SP^{S,s} \bigr)^*  =
  \SP^{S,s} \bigm|_{q\to \overline{q}} \, =
  \PP \, \SP^{S,s} \, \PP \,.
\end{equation}
In fact, by Remark~\ref{eta0}, it is evident that these
projectors are quasi--Hermitian operators.

Consider a one dimensional lattice which contains $N$ nodes,
each node carries an irreducible module $V^S$ as a local Hilbert
space. For an operator $\SA$ in $V^S$ or in $(V^S)^{\ot 2}$,
we will use the standard notations $\SA_{n}$ and $\SA_{nm}$
for its embedding in operators in
${\mathfrak H} \,{=}\,\bigl(V^S\bigr)^{\otimes \klein N}$
that act non--trivially only in the $n$--th or in the $n$--th
and $m$--th tensor components, respectively.
The following operator
\beq{Hal}
  \SH^{S,s}_{\{a_1,\ldots,a_{N-1}\}} =
    \sum_{n=1}^{N-1} a_n \, \SP^{S,s}_{n,n+1} \,,
    \qquad  a_n \in \mathbb R \,,
\end{equation}
can be regarded as the Hamiltonian of an open spin chain with
{\em inhomogeneous} coupling. This Hamiltonian commutes with the
global action of $U_q(sl_2)$ in~$\mathfrak H$, i.e.
we have (see Appendix~\ref{APP})
\beq{HUsym}
  \bigl[\, \SH^{S,s}_{\{a_1,\ldots,a_{N-1}\}} ,
  \pis^{\ot \klein N} \bigl( \Delta^{(N-1)} (X) \bigr)\,
  \bigr]  = 0 \,, \qquad
  \text{for any }\  X \,{\in}\, U_q(sl_2) \,.
\end{equation}
Here and in the rest of the text we use the abbreviation
$\pis^{\otimes \klein N} \,{\equiv}\,
 (\pis \,{\ot}\, \ldots \,{\ot}\, \pis)$.

Recall that the positive integer power of the
comultiplication used in \rf{HUsym} is defined
recursively: $\Delta^{(1)} \,{\equiv}\, \Delta$ and
$\Delta^{(N)} \,{=}\, \Delta_{N,n} \,{\circ} \, \Delta^{(N-1)}$.
Here and below we denote   $\Delta_{N,n}  \,{\equiv}\,
   id_{n-1} \,{\ot}\, \Delta \,{\ot}\, id_{N-n}$,
where $n$ can be taken any from $1$~to~$N$ thanks to the
coassociativity of $\Delta$, i.e.
$\Delta_{2,1} \,{\circ} \, \Delta \,{=}\,
\Delta_{2,2} \,{\circ} \, \Delta$.

\begin{rem}
The Hamiltonian \rf{Hal} is pseudo--Hermitian in the
{\em homogeneous} case ($a_1\,{=}\,{\ldots}\,{=}\,a_{N-1}$)
for any $N$ and in the {\em two--periodic} case
($a_{2n+1} \,{=}\, a_1$, $a_{2n} \,{=}\, a_2$) for even~$N$.
The symmetrizing operator for these cases is given by
$\eta \,{=}\, \PP_{1,N}\PP_{2,N-1} \ldots$.
\end{rem}

In general, a lattice model with Hamiltonian \rf{Hal} is not
integrable. However, its homogeneous case
is integrable for $s\,{=}\,0$.
The corresponding R--matrix is constructed by
a Baxterization of the Temperley--Lieb algebra (see, e.g.~\cite{Ku}).
In particular, for $S\,{=}\, \frac{1}{2}$ and $s\,{=}\,0$,
setting $a_1 \,{=}\, a_2 \,{=}\, {\ldots} \,{=}\, {-}\cos \gamma$,
we recover the Hamiltonian of the well known XXZ model of
spin~$\fr 12$ (which is an integrable deformation of
the Heisenberg chain),
\beq{Hxxz}
  \SH^{\frac 12,0}_{\{-\cos\gamma,\ldots\}}
 = \!\sum_{n=1}^{N-1} \Bigl(
 \fr{1}{2} (\sigma^+_n \sigma^-_{n+1} + \sigma^-_n \sigma^+_{n+1})
 + \fr{\cos\gamma}{4} \bigl(\sigma^\3_n \sigma^\3_{n+1} - 1 \bigr)
 + \fr{i \sin\gamma}{4} (\sigma^\3_n - \sigma^\3_{n+1}) \Bigr).
\end{equation}

\subsection{$N=2$ and $N=3$  }
We commence by studying spectra of short chains.
Since $\mathfrak H$ is finite dimensional, we have
$\mathrm{ Spec} \, \SH =
\bigl\{\lambda : {\cal P}_\SH(\la)\,{=}\,0 \bigr\}$,
where ${\cal P}_\SH(\la)$ is the minimal polynomial for $\SH$,
i.e. the least degree non--zero polynomial
such that ${\cal P}_\SH(\SH) \,{=}\,0$.
In the simplest case, $N\,{=}\,2$, we have
$\SH^{S,s}_{\{a_1\}}= a_1 \SP^{S,s}_{12}$.
The corresponding minimal polynomial is
${\cal P}^{S,s}_{a_1}(\lambda) \,{=}\, \la^2 \,{-}\,a_1 \lambda$,
which shows that the spectrum consists of points $0$ and $a_1$ and
thus is real.

For $N=3$, we have
$\SH^{S,s}_{\{a_1,a_2\}}= a_1 \SP^{S,s}_{12} + a_2 \SP^{S,s}_{23}$.
Let us consider first the case $s\,{=}\,0$.
In this case the projectors satisfy the
relations of the Temperley--Lieb algebra \cite{BB1,B2}:
\beq{TLa}
  \SP^{S,0}_{n-1,n} \, \SP^{S,0}_{n,n+1} \, \SP^{S,0}_{n-1,n}
 = \mus \, \SP^{S,0}_{n-1,n} \,, \qquad
 \mus = \fr{1}{[2S+1]^2} \,.
\end{equation}
Using these relations (see Appendix~\ref{AMP}), we find the minimal
polynomial for $\SH^{S,0}_{\{a_1,a_2\}}$:
\beq{mP03}
 {\cal P}^{S,0}_{a_1,a_2}(\lambda) =
    \la\, \bigl(\la^2 -(a_1 \,{+}\, a_2) \, \la +
 a_1 a_2 \, (1-\mus) \bigr) \,.
\end{equation}
Hence it follows that all eigenvalues of
$\SH^{S,0}_{\{a_1,a_2\}}$ are real iff
${\cal D}^{S,0}\equiv (a_1 \,{-}\, a_2)^2 + 4a_1 a_2 \mus$
is non--negative, that is iff
\beq{D03}
 \Bigl( \frac{\sin{(2S{+}1)\gamma}}{\sin{\gamma}} \Bigr)^2 \geq
    - \frac{4 a_1 a_2}{(a_1 - a_2)^2} \,.
\end{equation}
Clearly, this condition holds always if $a_1$ and $a_2$ are
both positive (or both negative). If $a_1 a_2 \,{<}\, 0$, then
the spectrum of $\SH^{S,0}_{\{a_1,a_2\}}$ is not real for those
values of~$\gamma$ where \rf{D03} does not hold.
Note that the r.h.s. of \rf{D03} attains
the maximal value equal to~1 when $a_2 \,{=}\, {-}a_1$.
Hence we infer that, even for $a_1 a_2 \,{<}\, 0$, the spectrum
of $\SH^{S,0}_{\{a_1,a_2\}}$ is guaranteed to be real for
sufficiently small values of $\gamma$, namely for
$|\gamma| \,{<}\, \gamma_{S,0}$, where
\beq{ga03}
 \gamma_{S,0} = \frac{\pi}{2(S\,{+}\,1)}
\end{equation}
is the minimal positive solution of the
equation $\sin{(2S{+}1)\gamma}=\sin\gamma$.

For $s \neq 0$, the projectors $\SP^{S,s}$ do not satisfy relations of
the type \rf{TLa}. However, by evaluating \rf{dC2} and \rf{PjC} in
the representation \rf{EFK}, one can find an explicit matrix form
of these projectors and then search for the coefficients of the
minimal polynomial for $\SH^{S,s}_{\{a_1,a_2\}}$.
The author performed these steps for $S\,{=}\,1,\frac{3}{2}$
and $s\,{\leq}\, 2S$ using
{\em Mathematica}$^{\klein \text{TM}}$.
The polynomials obtained are:
\begin{align}
\label{mP11}
 {\cal P}^{S,s}_{a_1,a_2}(\lambda) =
 \la^{\epsilon_{S,s}} \prod_k \bigl(\la^2 -(a_1 \,{+}\, a_2) \, \la +
 a_1 a_2 \, (1-d_k^{S,s}) \bigr) \,,
\end{align}
where the coefficients $d_k^{S,s}$ are listed in Appendix~\ref{CMP}.
In \rf{mP11} we have $\epsilon_{\klein S,s} \,{=}\,0$ if there is
$d_k^{S,s} \,{=}\,1$ in the list for given $S$ and $s$
(which occurs for $s\,{=}\,2S$)
and $\epsilon_{\klein S,s} \,{=}\,1$ otherwise.

{}From \rf{mP11} we infer that all eigenvalues of
$\SH^{S,s}_{\{a_1,a_2\}}$ are real iff all
${\cal D}^{S,s}_{k}\equiv (a_1 \,{-}\, a_2)^2 + 4a_1 a_2  d_k^{S,s}$
are non--negative, that is iff
\beq{DsS}
 \bigl( d_k^{S,s} \bigr)^{-1}   \geq
    - \frac{4 a_1 a_2}{(a_1 - a_2)^2} \,.
\end{equation}
Thus, we see that, for the considered values of $S$,
the spectrum of $\SH^{S,s}_{\{a_1,a_2\}}$ is real if $a_1 a_2 \,{>}\,0$
and is not real for some values of~$\gamma$ if $a_1 a_2 \,{<}\,0$.
In the latter case, the spectrum of $\SH^{S,s}_{\{a_1,a_2\}}$ is
guaranteed to be real for
$|\gamma| < \gamma_{S,s} \,{=}\, \min_k \gamma_{S,s}^{\klein \{k\}}$,
where $\gamma_{S,s}^{\klein \{k\}}$ is the minimal positive
solution of the equation $ d_k^{S,s} =1$.
In Appendix~\ref{CMP}, the coefficients $ d_k^{S,s}$ are listed
in such a way that $k\,{=}\,1$ corresponds to the minimal value
among $\gamma_{S,s}^{\klein \{k\}}$. The list \rf{ga1} of resulting
values $\gamma_{S,s}$ together with formula \rf{ga03} allows
us to conjecture the following.
\begin{conj}
For $a_1 a_2 \,{<}\,0$, the spectrum of $\SH^{S,s}_{\{a_1,a_2\}}$
is real for $|\gamma| \,{<}\, \gamma_{S,s}$, where
\beq{ga05}
 \gamma_{S,s}=
 \frac{\pi}{2(s \,{+}\, S\,{+}\,1 \,{-}\,\delta_{s,2S})} \,.
\end{equation}
\end{conj}
\begin{rem}
Appearance of the correction for $s \,{=}\, 2S$ in \rf{ga05}
seems to be related to the fact that
$\SP^{S,2S} \,{=}\, {\sf 1} - \sum_{s\neq 2S} \SP^{S,s}$.
In particular, \rf{ga05} yields
$\gamma_{\frac 12,1}=\gamma_{\frac 12,0}$, as should be
anticipated because
$\SH^{{\klein \frac 12},1}_{\{a_1,a_2\}}$ and
$\SH^{{\klein \frac 12},0}_{\{a_1,a_2\}}$
differ only by a sign and a shift by a real multiple
of the identity operator.
\end{rem}

\subsection{$N=4$ and $N=5$ for $s=0$ }

For $N=4$ and $s\,{=}\,0$, a computation analogous to that
in Appendix~\ref{AMP} yields the following minimal polynomial
\begin{align}\label{mP04}
 {\cal P}^{S,0}_{a_1,a_2,a_3}(\lambda) =& \,
    \la \, \bigl(\la^2 -(a_1 \,{+}\, a_2 \,{+}\, a_3) \, \la +
 (a_1 \,{+}\, a_3) \, a_2 \, (1 \,{-}\, \mus) \bigr) \\
\nonumber
{} \times & \bigl(\la^3 -(a_1 \,{+}\, a_2 \,{+}\, a_3)  \la^2 +
 \bigl(a_1 a_3 +a_2 (a_1 \,{+}\, a_3) (1 \,{-}\, \mus) \bigr)  \la
 - a_1 a_2 a_3 \, (1-2\mus) \bigr) \,.
\end{align}
Analysis of the reality of the roots of the cubic factor is fairly
complicated. Therefore, we restrict our consideration to the
case $a_3\,{=}\,a_1$ (which, in particular, includes
the homogeneous case).
In this case, \rf{mP04} simplifies and acquires the following form:
\begin{align}\label{mP04b}
  {\cal P}^{S,0}_{a_1,a_2,a_1}(\lambda) =& \,
    \la \, (\la \,{-}\, a_1)
  \bigl(\la^2 -(a_1 \,{+}\, a_2) \, \la +
  a_1 a_2 \, (1 \,{-}\, 2\mus) \bigr) \\
  \nonumber
 {} & \times  \bigl(\la^2 -(2a_1 \,{+}\, a_2 ) \, \la +
  2 a_1 \, a_2 \, (1 \,{-}\, \mus) \bigr)   \,.
\end{align}
It follows from \rf{mP04b} that all eigenvalues of
$\SH^{S,0}_{\{a_1,a_2,a_1\}}$ are real iff both
$\tilde{\cal D}^{S,0}_{1} \,{\equiv}\, (2a_1 \,{-}\, a_2)^2
  + 8a_1 a_2 \mus$
and
$\tilde{\cal D}^{S,0}_{2} \,{\equiv}\, (a_1 \,{-}\, a_2)^2
  \,{+}\, 8a_1 a_2 \mus$
are non--negative.
Thus, we conclude that the spectrum of $\SH^{S,0}_{\{a_1,a_2,a_1\}}$
is real if $a_1 a_2 \,{>}\,0$ and is not real for some values
of~$\gamma$ if $a_1 a_2 \,{<}\,0$. In the latter case, we note
that $\tilde{\cal D}^{S,0}_{1} \,{-}\, \tilde{\cal D}^{S,0}_{2}
= a_1(3a_1\,{-}\,2a_2) \,{>}\, 0$. Therefore, for
$a_1 a_2 \,{<}\,0$, the spectrum of
$\SH^{S,0}_{\{a_1,a_2,a_1\}}$  is  real iff
$\tilde{\cal D}^{S,0}_{2} \,{>}\, 0$, that is iff
\beq{D04}
 \Bigl( \frac{\sin{(2S{+}1)\gamma}}{\sin{\gamma}} \Bigr)^2 \geq
    - \frac{8 a_1 a_2}{(a_1 - a_2)^2} \,.
\end{equation}
The r.h.s. of \rf{D04} attains the maximal value equal to~2 when
$a_2 \,{=}\, {-}a_1$. Thus, for $a_1 a_2 \,{<}\, 0$, the spectrum
of $\SH^{S,0}_{\{a_1,a_2,a_1\}}$ is guaranteed to be real for
$|\gamma| < \tilde\gamma_{S,0}$, where $\tilde\gamma_{S,0}$
is the minimal positive solution of the
equation $\sin^2{(2S{+}1)\gamma}=2\sin^2\gamma$.
Taking into account that, for $S\,{\geq}\,\frac{1}{2}$, we have
$\sin{(2S{+}1)\gamma}/\sin\gamma \,{>}\, \sqrt{2}$
on some interval that contains the point $\gamma\,{=}\,0$,
the value $\tilde\gamma_{S,0}$ can be equivalently determined
as the minimal positive solution of the equation
\beq{U2S}
U_{2S}(\cos\gamma)=\sqrt{2} \,,
\end{equation}
where $U_n(t)$ is the Chebyshev polynomial of the second
kind ($U_1(t)\,{=}\,2t$, $U_2(t)\,{=}\,4t^2 \,{-}\,1$, etc.)
In particular, we have
\beq{tgaS}
\tilde\gamma_{\frac{1}{2},0}=\frac{\pi}{4} \,, \qquad
\tilde\gamma_{1,0}=\arccos\fr{\sqrt{1+\sqrt{2}}}{2}
 \approx 0.217 \, \pi \,.
\end{equation}

For $N=5$ and $s\,{=}\,0$, even in the reduced case $a_3\,{=}\,a_1$,
$a_4\,{=}\,a_2$, the minimal polynomial
${\cal P}^{S,0}_{a_1,a_2,a_1,a_2}(\lambda)$ contains
factors which are fourth and fifth degree polynomials in~$\la$.
However, for $a_1\,{=}\,a_3\,{=}\,a$,\ $a_2 \,{=}\,a_4\,{=}\,{-}a$,
it simplifies and acquires the following form
\begin{align}\label{mP05b}
 {\cal P}^{S,0}_{a,-a,a,-a}(\lambda) =& \,
    \la \, \bigl(\la^4 + a^2 (3 \mus \,{-}\, 2) \, \la^2 +
  a^4 \, (\mu_S^2 \,{-}\, 3\mus \,{+}\, 1) \bigr) \\
  \nonumber
 {} & \times  \bigl(\la^4 + a^2 (6 \mus \,{-}\, 5) \, \la^2 +
  a^4 \, (5 \mu_S^2 \,{-}\, 10\mus \,{+}\, 4) \bigr)   \,.
\end{align}
The first bi--quadratic factor here has only real roots
iff $\mus \,{\leq}\, \frac{3-\sqrt{5}}{2}$. For this
range of $\mus$, the second bi--quadratic factor has also
only real roots. Thus, the spectrum of
$\SH^{S,0}_{\{a,-a,a,-a\}}$ is guaranteed to be real for
$|\gamma| < \tilde\gamma_{S,0}$, where $\tilde\gamma_{S,0}$
is the minimal positive solution of the equation
$\sin{(2S{+}1)\gamma}=\bigl(\frac{3+\sqrt{5}}{2}\bigr)^{1/2}\sin\gamma$,
or, equivalently, of the equation
\beq{U2Sb}
U_{2S}(\cos\gamma)=\frac{1+\sqrt{5}}{2} \,.
\end{equation}
In particular, we have
\beq{tgaSb}
\tilde\gamma_{\frac{1}{2},0}=\tilde\gamma_{1,0}
    =\frac{\pi}{5} \,, \qquad
\tilde\gamma_{\frac{3}{2},0}  \approx 0.172 \, \pi \,.
\end{equation}

Equations \rf{ga03}, \rf{U2S}, and \rf{U2Sb} allow us to
make the following conjecture about a chain 
with {\em alternating} coupling
($a_1 \,{=}\, {-} a_2 \,{=}\, a_3 \,{=}\, {-} a_4 \,{=}\,{\ldots}$).

\begin{conj}\label{Conj2}
For an alternating chain with $N {\geq}\,3$ nodes, the spectrum of
$\SH^{S,0}_{\{a,-a,a,-a,\ldots\}}$ is real for
$|\gamma| < \tilde\gamma_{S,0}$, where $\tilde\gamma_{S,0}$
is the minimal positive solution of the equation
\beq{U2SN}
 U_{2S}(\cos\gamma)= 2 \cos \frac{\pi}{N} \,.
\end{equation}
\end{conj}
\begin{rem}
For the alternating chain of spin $S\,{=}\,\fr 12$ and length $N$,
Eq.~\rf{U2SN} yields
\beq{altN}
 \tilde\gamma_{\frac{1}{2},0} =\frac{\pi}{N} \,,
\end{equation}
which is the most natural extrapolation of the values
$\tilde\gamma_{\frac{1}{2},0}$ given by
Eqs. \rf{ga03}, \rf{tgaS}, and~\rf{tgaSb}.
\end{rem}

\subsection{A universal metric operator}
The most general form of a $U_q(sl_2))$--invariant open spin
chain Hamiltonian with a nearest--neighbour interaction
and an inhomogeneous coupling is the following
\beq{Hal2}
  \SH^{S}_{N} =
    \sum_{n=1}^{N-1} \sum_{s=0}^{2S}
    b_{n,s} \, \SP^{S,s}_{n,n+1}  \,,
    \qquad  b_{n,s} \in \mathbb R \,.
\end{equation}
The previously considered Hamiltonian \rf{Hal} is a particular
case of \rf{Hal2} corresponding to the choice
$b_{n,s'} \,{=}\, a_n \delta_{ss'} $. A particular homogeneous
case of \rf{Hal2} corresponding to the choice
$b_{n,s} \,{=}\, (\sin\gamma)\sum_{k=1}^s \cot(\gamma k)$
recovers the Hamiltonian of the integrable XXZ model
of spin~$S$ (see e.g.~\cite{B1}).
For spin $S \,{=}\,1$, another integrable model recovered as
a homogeneous case of \rf{Hal2} is the spin chain generated
by the Izergin--Korepin R--matrix~\cite{IK}.

Now our aim is to construct a {\em universal} metric operator
$\eta_{\klein N}$ for the Hamiltonian~\rf{Hal2}, i.e. such that
relation \rf{Heta} holds irrespective of the choice of the
coupling coefficients~$b_{n,s}$. As seen from Eq.~\rf{P*},
it suffices to find such $\eta_{\klein N}$ that the relation
\beq{etaP}
  \eta_{\klein N} \, \SP^{S,s}_{n,n+1} =
 \bigl( \SP^{S,s}_{n,n+1} \bigr)^* \, \eta_{\klein N} =
  \SP^{S,s}_{n+1,n} \, \eta_{\klein N}
\end{equation}
holds for all $n=1,\ldots,N\,{-}\,1$.

Recall that the Hopf algebra $U_q(sl_2)$ is
quasi--triangular~\cite{D1}, i.e. it possesses a universal
R--matrix which is an invertible element of (a completion of)
$U_q(sl_2)^{\ot 2}$ with the following properties
\begin{eqnarray}
\label{uniRa}
 & R \, \Delta (X) = \Delta' (X) \, R \,, \qquad
 \text{ for any }\ X \in U_q(sl_2) \,, & \\
\label{uniRb}
 & (\Delta \,{\ot}\, id) \, R = R_{13} \, R_{23} \,, \qquad
 (id \,{\ot}\, \Delta) \, R = R_{13} \, R_{12} \,, &
\end{eqnarray}
where $\Delta' (X) \equiv \PP \, \Delta (X) \, \PP$.
In fact, there exist two universal R--matrices because,
if $R^+ \,{\equiv}\, R$ satisfies \rf{uniRa}--\rf{uniRb},
then so does $R^- = \PP \bigl( R^+ \bigr)^{-1} \PP$.
The explicit form of the universal R--matrices  $R^\pm$
consistent with  the comultiplication \rf{DelEFK}
is given in Appendix~\ref{AUR}.

Let us denote  $\SR^\pm \equiv (\pis \ot \pis) R^\pm$.
Eq.~\rf{uniRa} along with the fact that $\SP^{S,s}$
is a function of $(\pis \ot \pis)\Delta(C)$
(see~Eq.~\rf{PjC}) implies that the projectors $\SP^{S,s}$
are symmetrizable by~$\SR^\pm$, i.e.
\beq{RPb}
   \SR^\pm_{n,n+1} \, \SP^{S,s}_{n,n+1} =
  \SP^{S,s}_{n+1,n} \, \SR^\pm_{n,n+1} \,.
\end{equation}
Eq.~\rf{R*} implies that
$\eta^{\klein S}_\2(\alpha) \,{=}\,
e^{i\alpha} \, \SR^+ \,{+}\,  e^{-i\alpha} \, \SR^{-}$
is a Hermitian operator if $\alpha \,{\in}\,\mathbb R$.
This, along with \rf{RPb}, means that
$\eta^{\klein S}_\2(\alpha)$ is a one--parametric
family of symmetrizing operators for a chain of length
$N \,{=}\, 2$.
We will extend this observation to a chain of arbitrary
length as follows (a proof is given in Appendix~\ref{AP2}).

\begin{propn}\label{PEN}
a) For a chain of length $N$, the following
operators satisfy relations~\rf{etaP}
\beq{etaN}
  \eta^\pm_{\klein N} =
  \stackrel{\leftarrow}{\SR}_{N} {\ldots}
 \stackrel{\leftarrow}{\SR}_{2}, \qquad \text{where}\quad
 \stackrel{\leftarrow}{\SR}_n = \SR_{n-1,n} \ldots \SR_{1,n} \,.
\end{equation}
b) These operators can also be represented as follows
\beq{etaNop}
 \eta^\pm_{\klein N} =
 \stackrel{\rightarrow}{\SR}_1 {\ldots}
 \stackrel{\rightarrow}{\SR}_{N-1}, \qquad \text{where}\quad
 \stackrel{\rightarrow}{\SR}_n = \SR_{n,n+1} \ldots \SR_{n,N} \,.
\end{equation}
c) These operators are conjugate to each other,
\beq{eta*}
   \bigl( \eta^+_{\klein N} \bigr)^* = \eta^-_{\klein N} \,.
\end{equation}
\end{propn}

\begin{rem}
The proof of Proposition~\ref{PEN} is facilitated
by an observation that the operation
$\Delta^\pm \,{\equiv}\, R^\pm \Delta$ is
coassociative (but note that it is not an algebra homomorphism)
and that the operators \rf{etaN} can be expressed
in terms of its power:
$\eta^\pm_{\klein N} \,{=}\, \pis^{\ot \klein N}
 \bigl( \Delta_\pm^{(N-1)} \, (1) \bigr)$,
see Lemma~2.
\end{rem}

As seen from \rf{eta*}, the symmetrizing operators
$\eta^\pm_{\klein N}$ are not Hermitian. However,
we can utilize them to build a multi--parametric family of
Hermitian symmetrizing operators as follows:
\beq{etaal2}
    \eta_{\klein N}^{\klein S} (\alpha_1,{\ldots}|\beta_1,{\ldots}) =
  \sum_{n \geq 1} \beta_n \bigl(
 e^{i\alpha_n} \, \eta^+_{\klein N}
 \bigl( (\eta^-_{\klein N})^{-1} \, \eta^+_{\klein N} \bigr)^{n-1}
 +
 e^{-i\alpha_n} \, \eta^-_{\klein N}
 \bigl( (\eta^+_{\klein N})^{-1} \, \eta^-_{\klein N} \bigr)^{n-1} \bigr) ,
\end{equation}
where all $\alpha_n$ and $\beta_n$ are real.
Here we used a simple fact: if $\eta$, $\eta'$, and $\eta''$
are symmetrizing operators for an operator $\SH$, then so is
$\eta (\eta')^{-1} \eta''$ if $\eta'$ is invertible.
In our case, $\eta^\pm_{\klein N}$ are invertible because
so are the universal R--matrices.

Note that, for 
$\gamma\,{=}\,0$, we have
$\SR^\pm \,{=}\,{\sf 1}\,{\ot}\,{\sf 1}$
and $\eta^\pm_{\klein N} \,{=}\,{\sf 1}_{\klein N}$. Therefore,
for sufficiently small values of $\gamma$ and appropriately
chosen coefficients $\{\alpha_n\}$, $\{\beta_n\}$,
operator \rf{etaal2} is positive definite and, thus,
is a metric operator for the Hamiltonian~\rf{Hal2}.

For $\gamma\,{\neq}\,0$, it is not straightforward to determine
the values of $\{\alpha_n\}$ and $\{\beta_n\}$ for which
\rf{etaal2} is positive definite. In the present article,
we restrict our consideration to a one--parametric family,
\beq{etaal}
    \eta_{\klein N}^{\klein S} (\alpha)  =
 e^{i\alpha} \, \eta^+_{\klein N} +
 e^{-i\alpha} \, \eta^-_{\klein N} \,, \qquad
 \alpha \in {\mathbb R} \,.
\end{equation}
Let $\gamma(\alpha)$ denote
the maximal positive value of $\gamma$ for which \rf{etaal}
is positive definite for given~$\alpha$, and let
$\hat{\gamma}_{\klein S} \,{\equiv}\,
 \sup_{\alpha} \gamma(\alpha)$.
At least one of the eigenvalues of
$\eta_{\klein N}^{\klein S} (\alpha)$ vanishes at
$\gamma \,{=}\, \hat{\gamma}_{\klein S}$. Therefore,
$\hat{\gamma}_{\klein S}$ can be determined from
the condition
$\det\bigl( \eta_{\klein N}^{\klein S} (\alpha)\bigr) \,{=}\,0$.

\begin{lem}\label{DE}
The following relation holds
\beq{specRN}
   \det\bigl( \eta_{\klein N}^{\klein S} (\alpha) \bigr) =
 \prod_{s=s_\0}^{SN} \Bigl(e^{i\alpha} q^{s(s+1)-NS(S+1)} +
 e^{-i\alpha} q^{NS(S+1) -s(s+1)} \Bigr)^{(2s+1)\nu_s } \,,
\end{equation}
where $\nu_s$ are the multiplicities of the irreducible submodules
in the decomposition 
$\bigl(V^{S}\bigr)^{\ot \klein N}
 \,{=}\, \mathop{\oplus}\limits_{s= s_\0}^{NS} \nu_s V^s$. Here
$s_\0 \,{=}\, 0$ if $NS$ is integer and
$s_\0 \,{=}\, \frac 12$ if $NS$ is half--integer.
\end{lem}


The range of $\gamma$ that includes the point $\gamma \,{=}\, 0$
and in which \rf{specRN} does not vanish is maximal if
we set $\alpha \,{=}\, \alpha_0 \equiv
\frac{\gamma}{2}
\bigl(NS(2S \,{+}\, 1 \,{-}\,NS) \,{-}\, s_\0(s_\0 \,{+}\, 1)\bigr)$.
Then we have
$\det\bigl( \eta_{\klein N}^{\klein S} (\alpha_0) \bigr) \,{>}\,0$
for  $|\gamma| \,{<}\, \hat{\gamma}_{\klein S}$, where
\beq{gaN}
  \hat{\gamma}_{\klein S} =
 \frac{\pi}{(NS \,{-}\,s_\0)(NS \,{+}\, s_\0 \,{+}\, 1)}  \,.
\end{equation}
Since $\fr{1}{2}\eta_{\klein N}^{\klein S} (0) \,{=}\, {\sf 1}$
for $\gamma \,{=}\,0$, we conclude that
$\eta_{\klein N}^{\klein S} (\alpha_0)$ is positive
definite for $|\gamma| \,{<}\, \hat{\gamma}_{\klein S}$.
Thus, we have established the following.


\begin{propn}\label{PHQH}
The Hamiltonian $\SH^{S}_{N}$ given by \rf{Hal2}
is quasi--Hermitian for any choice of the coupling
constants $b_{n,s}$ provided that
$|\gamma| \,{<}\, \hat{\gamma}_{\klein S}$,
where $\hat{\gamma}_{\klein S}$ is given by~\rf{gaN}.
\end{propn}

\section*{Conclusion}

It is well known that for a given quasi--Hermitian operator $\SH$
there are many metric operators~\cite{SGH,Be,M2}.
In the physical literature on non--Hermitian Hamiltonians,
the one most frequently discussed is the operator $\eta_{\klein 0}$
considered in Remark~\ref{eta0}. For the case of $\SH$ having
a simple spectrum, a generalization of $\eta_{\klein 0}$
to an operator of the type \rf{etappdeg} was given in~\cite{ZG}.
In the present article, we have given the most general form of a
metric operator for a finite dimensional quasi--Hermitian operator
$\SH$ not assuming its spectrum to be simple.

As an example of a compound operator \rf{HN} given by the sum of
quasi--Hermitian operators, we studied the Hamiltonians \rf{Hal}
and \rf{Hal2} of an open $U_q(sl_2)$--invariant spin chain of
spin $S$ and length~$N$. For these Hamiltonians, we constructed
two symmetrizing operators $\eta^\pm_{\klein N}$ in terms of
products of local R--matrices (let us note that similar products
appeared in a different context in~\cite{TV}). From the operators
$\eta^\pm_{\klein N}$ we built a multi--parametric family of
metric operators.
These metric operators are universal, i.e. independent of the
coupling constants, and thus non--dynamical, i.e. their
construction does not require the knowledge of the eigenvectors
of a Hamiltonian.

By optimizing the value of the free parameter in
a one--parametric subfamily of universal metric operators,
we obtained an estimate \rf{gaN}
on the range  of the deformation parameter $\gamma$ in which
the considered Hamiltonians are quasi--Hermitian.
Note that this range is in general narrower than the ranges of
$\gamma$ for which the short chains considered in Section~2.2 and 2.3
have real spectra.
We expect that better estimates of the quasi--Hermiticity range
can be obtained by using the multi--parametric family~\rf{etaal2}.

It is worth mention that the most general family  \rf{Hal2} of
Hamiltonians includes, in particular, all known (see, e.g. \cite{B2})
integrable $U_q(sl_2)$--invariant spin chains with nearest--neighbour
interaction: the XXZ model of spin $S$,
the Temperley--Lieb spin chain of spin $S$, and, for spin $1$,
the  spin chain generated by the Izergin--Korepin
R--matrix. So our construction of the metric operators
applies also to these cases.

Let us conclude with several remarks on the ``experimental''
data obtained in Section~2.2 and 2.3 for the ranges
of $\gamma$ in which the Hamiltonian \rf{Hal} has a real spectrum.
First, it is very interesting to note that the value
of $\tilde{\gamma}_{{\klein\frac 12},0}$ in \rf{altN} for an
alternating XXZ chain of spin $\fr 12$ is exactly the same as the
boundary of the quasi--Hermiticity range for
a homogeneous XXZ chain of spin~$\fr 12$ found in \cite{KW}
by means of the path basis technique.
Actually, the results for short chains seem to indicate
that, for given $S$ and $N$, the alternating chain
($a_1 \,{=}\, -a_2 \,{=}\, a_3  \,{=}\, -a_4 {\ldots}$) is
the most non--Hermitian one, at least in the subclass of
chains with a two--periodic coupling
($a_{2n+1} \,{=}\, a_1$, $a_{2n} \,{=}\, a_2$).
Thus, we have a reason to expect that
Conjecture~\ref{Conj2} may hold not only for alternating
but also for two--periodic chains and, possibly, even for
arbitrary ones.

Finally, let us remind that
in the general $N \,{=}\,3$ case and the two--periodic
$N \,{=}\,4$ case the spectra are always real if all
coupling constants are positive. This observation
is supported by numerical checks in a number of other cases.
It is thus tempting to suggest the following.

\enlargethispage{1\baselineskip}
\begin{conj}
For $|\gamma| < \frac{\pi}{2S}$, the Hamiltonian \rf{Hal}
of a spin chain with inhomogeneous
coupling has a real spectrum if all $a_n >0$.
\end{conj}

\appendix
\section{Appendix}
\subsection{Proof of Proposition~\ref{etadeg} }\label{AP1}

The spectral resolutions of a quasi--Hermitian operator
$\SH$ and its adjoint are
$ \SH \,{=}\, \sum_{j= 1}^{d'} \lambda_j  \mathfrak{P}_j $,
$ \SH^*  \,{=}\, \sum_{j= 1}^{d'} \lambda_j  \mathfrak{P}^*_j $,
where
$\mathfrak{P}_j \,{=}\, \sum_{k=1}^{\mu_j}  \SP_{j,k}$
are the projectors onto the subspaces~$\mathfrak H_j$.
Hence
\beq{pfunh}
\mathfrak{P}_j=\prod_{n \neq j}^{d'}
    \frac{\SH - \la_n{\sf 1}}{\la_j - \la_n} \,,\qquad
\mathfrak{P}^*_j=\prod_{n \neq j}^{d'}
    \frac{\SH^* - \la_n{\sf 1}}{\la_j - \la_n} \,.
\end{equation}

It follows from relation~\rf{Heta} that
$\eta\, \SH^n \,{=}\, (\SH^*)^n \eta$ for all $n \,{\in}\, \mathbb N$.
Therefore $\eta f(\SH) \,{=}\, \bigl(f(\SH)\bigr)^*\eta$,
where $f(t)$ is an arbitrary
polynomial with real coefficients. Along with \rf{pfunh} it implies that
a positive definite operator $\eta$ is a metric operator for $\SH$ iff
\beq{etapp*}
    \eta \, \mathfrak{P}_j = \mathfrak{P}^*_j \, \eta \,,
 \qquad j=1,\ldots,d'.
\end{equation}
As the basis of $\mathfrak{H}$ we take a naturally ordered set
$\{\om_{1,1},{\ldots},\om_{1,\mu_1},\om_{2,1},
    {\ldots},\om_{d',\mu_{d'}}\}$.
Then, according to \rf{PPj}, we have
$\tilde{O}(\mathfrak{P}_j)=G \, E_j$
and $\tilde{O}(\mathfrak{P}^*_j)= E_j G$, where $E_j$ is a diagonal
matrix with $\mu_j$ consecutive entries equal to~1 and others
being~0; the identity matrix has the resolution
$E=\sum_{j=1}^{d'} E_j$.
Using \rf{OOt1}, we find that
$\tilde{O}(\eta\, \mathfrak{P}_j)=\tilde{O}(\eta) E_j$ and
$\tilde{O}(\mathfrak{P}^*_j \eta)=E_j \tilde{O}(\eta)$.
Therefore, \rf{etapp*} holds iff $\tilde{O}(\eta)$ commutes with
$E_j$ for all~$j$, that is iff $\tilde{O}(\eta)$ is a
block diagonal matrix. The second relation in \rf{OOt2}
implies that $O(\eta^{-1})$ is inverse to $\tilde{O}(\eta)$
and so it is also a block diagonal matrix.
Whence Eqs.~\rf{etappdeg} follow.
The Hermiticity of $\eta$ is equivalent to
$(\tilde{O}(\eta))^* \,{=}\, (\tilde{O}(\eta))$ which implies that
blocks $\Phi_j$ in \rf{etappdeg} must be Hermitian. Since $\eta$
is invertible, it is positive definite whenever $\eta^{-1}$ is so.
The latter condition requires, in particular, that
$\scp{x_j}{\eta^{-1} x_j} \,{>}\,0$, for
any non--zero vector $x_j \,{\in}\, {\mathfrak H}_j$.
Which is equivalent to
$\sum_{k,n= 1}^{\mu_j}
\bigl(\Phi_j^{-1}\bigr)_{kn} \, \overline{\beta_k} \beta_n  \,{>}\,0$,
where $\beta_k \,{\equiv}\, \scp{\om_{j,k}}{x_j}$ can be arbitrary
(but not all zero).
Thus, $\Phi_j^{-1}$ must be positive definite,
and hence so does~$\Phi_j$.

To prove the part b), we fix some bases $\{\om^\0_{j,k}\}$ of
subspaces $\mathfrak{H}_j$. Consider $\eta$ and $\eta^{-1}$
given by \rf{etappdeg} with some matrices~$\Phi^\0_j$.
Let $U_j$ be such unitary matrices that
$\Phi_j = U_j \Phi^\0_j U_j^{-1}$ are diagonal. Then,
introducing new basis vectors,
$\om_{j,k} = \sum_n (U^{-1}_j)_{kn} \om^\0_{j,n}$,
we achieve that, in the new basis, the symbol $O(\eta^{-1})$
becomes a diagonal matrix. The second relation in
\rf{OOt2} implies that $\tilde{O}(\eta)$ also becomes
a diagonal matrix. It remains to use formulae \rf{PPj}
to obtain Eqs.~\rf{etapp}.

\subsection{Projectors $\SP^{S,s}$ }\label{APP}

Let $q=e^{i\gamma}$.
The algebra \rf{Uq} has the following Casimir element:
\begin{equation}\label{C}
 C = \fr{1}{2} \bigl( E\,F + F\,E \bigr) -
    \fr{\cos\gamma}{4 \sin^2\gamma} \bigl( K - K^{-1}\bigr)^2 \,.
\end{equation}
Its value in an irreducible representation $V^S$ is
$\pis (C) = [S][S\,{+}\,1]$, where the
$q$--numbers are defined as
$[t] \equiv \fr{\sin \gamma t}{\sin \gamma}$.
The tensor Casimir element is an operator in $V^S \,{\ot}\, V^S$
given by
\begin{eqnarray}\label{dC2}
 & \SC^{S,S} =  (\pis \,{\ot}\, \pis) \Delta(C) =
(\pis \,{\ot}\, \pis) \Bigl( (K \, E) \otimes
 (F \, K^{-1}) + (F \, K^{-1}) \otimes (K \, E)  & \\
 & + \nonumber
 \frac{1}{2\sin^2\gamma} \bigl(
 ({\sf 1}\otimes {\sf 1} + K^2 \otimes  K^{-2})\, \cos\gamma -
 ({\sf 1} \otimes K^{-2} + K^{2} \otimes {\sf 1} )\,
\cos\bigl(\gamma(2S+1)\bigr) \bigr) \Bigr) . &
\end{eqnarray}
Obviously, we have $[\SC^{S,S} ,
 (\pis \,{\ot}\,\pis)\bigl(\Delta(X)\bigr) ]\,{=}\, 0$
for any $X\,{\in}\,U_q(sl_2)$.
Furthermore, we have
\beq{CX}
  [\SC^{S,S}_{n,n+1} ,
  \pis^{\ot \klein N} \bigl(\Delta^{(N-1)}(X)\bigr) ] =0 \,,
\end{equation}
for any $X$ and $n\,{=}\,1,{\ldots},N\,{-}\,1$.
This can be verified by evaluating
$\pis^{\ot \klein N} \bigl(\Delta_{N-1,n}(Y)\bigr)$, where
$Y \,{=}\, [C_n,\bigl(\Delta^{(N-2)}(X)\bigr) ] \,{=}\,0$.

With respect to the involution \rf{EFK*}, the tensor Casimir
element is not Hermitian but is a symmetrizable operator,
\beq{C*}
 \bigl( \SC^{S,S} \bigr)^*  =  \SC^{S,S}_{q^{-1}} =
 \PP\, \SC^{S,S}\, \PP  \,.
\end{equation}
Here $\SC^{S,S}_{q^{-1}}$ is the tensor Casimir element of
the algebra $U_{q^{-1}} (sl_2)$ (which is obtained by the mapping
$E\,{\to}\,E$, $F\,{\to}\,F$, $K\,{\to}\,K^{-1}$,
$q\,{\to}\,q^{-1}$).

The projectors $\SP^{S,s}$ can be constructed as follows
(see e.g. \cite{B1})
\beq{PjC}
 \SP^{S,s}  = \prod_{ \genfrac{}{}{0pt}{}{l=0}{l\neq s} }^{2S} \,
 \frac{\SC^{S,S} -[l] [l+1]}{ [s-l] [s+l+1]} \,.
\end{equation}
In particular, for $S \,{=}\, \fr 12$ we have
\begin{equation}\nonumber
 \SP^{\klein \frac 12,0} \,{=}\, \fr{1}{\kappa}
 \left(\begin{smallmatrix} 0 &&& \\
 & q^{-1} & -1 & \\  & -1 & q &  \\ &&& 0 \end{smallmatrix}\right) ,
 \qquad \SP^{\klein \frac 12,1} = \fr{1}{\kappa}
 \left(\begin{smallmatrix} \kappa &&& \\
 & q & 1 & \\  & 1 & q^{-1} &  \\ &&& \kappa \end{smallmatrix}\right) ,
 \qquad \kappa \,{=}\, q \,{+}\, q^{-1} .
\end{equation}

Note that matrix entries of $\SP^{S,s}$ can have singularities
at some values of~$\gamma$. This means that at these points
the Gram matrix of the basis of $V^S \,{\ot}\, V^S$ is not invertible
(cf. Eq.~\rf{PPj}) and some basis vectors become linear dependent.
We shall exclude such values of $\gamma$ from consideration.

Since $\SP^{S,s}$ are polynomials (with real coefficients)
in $\SC^{S,S}$, they satisfy the same relations \rf{CX}
and~\rf{C*}, i.e.,
\beq{PX}
 [\SP^{S,s}_{n,n+1} ,
  \pis^{\ot \klein N} \bigl(\Delta^{(N-1)}(X)\bigr) ] =0 \,,\qquad
  \bigl( \SP^{S,s} \bigr)^*  =  \SP^{S,s}_{q^{-1}} =
 \PP\, \SP^{S,s}\, \PP  \,.
\end{equation}
The first equality in the second relation implies, in particular,
that $\tilde{\om}_{s,k} \,{\simeq}\, \om_{s,k}|_{q\to \bar{q}}
 \,{=}\, \overline{\om}_{s,k}$,
where $\simeq$ means equality up to a normalization
(recall that $\tilde{\om}$ are, in general, not normalized,
cf.~Remark~\ref{omnorm}).
Using this relation and formulae \rf{PPj}, we can write down
a more explicit expression for $\SP^{S,s}$,
\beq{PPjq}
  \SP^{S,s} = \sum_{k=-s}^s \SP_{s,k}
 = \sum_{k=-s}^s \fr{1}{\kappa_{s,k}} \, \om_{s,k} \,
 \overline{\om}_{s,k}^\dagger \,,
\end{equation}
where $\kappa_{s,k} \,{=}\, \scp{\overline{\om}_{s,k}}{\om_{s,k}} \,{=}\,
 ||\om_{s,k}||^2_{\klein q\in\mathbb R}$, which is the norm of
$\om_{s,k}$ for $q \,{\in}\, \mathbb R$.
Consider, for instance, the case of $s\,{=}\,0$. The
corresponding submodule $V^0$ is one dimensional and it is easy to find
its basis vector $\om_{\klein 0,0}$ (which is annihilated by both
 $(\pis \,{\ot}\, \pis)\Delta(E)$ and $(\pis \,{\ot}\, \pis)\Delta(F)$),
\beq{00}
 \om_{\klein 0,0} = \sum_{k=-S}^S
 \frac{(-1)^{S-k} \, q^{-k}}{\sqrt{2S +1}} \,
 \om_{k} \ot \om_{-k}  \,,
\end{equation}
so that $\kappa_{\klein 0,0} \,{=}\, \fr{[2S +1]}{2S +1}$.
Substituting $\om_{\klein 0,0}$ in \rf{PPjq} and identifying
$\om_k \,{\simeq}\,e_{S+1-k}$, where $e_k$ is a vector in
${\mathbb C}^{2S+1}$ such that $(e_k)_r \,{=}\,\delta_{kr}$,
we obtain the following matrix form of $\SP^{S,0}$,
\beq{P00}
 \SP^{S,0} = 
 \sum_{m,n=1}^{2S+1}  \frac{(-1)^{m+n} \, q^{m+n-2S-2}}{[2S+1]} \,
 E_{m,n} \ot E_{2S+2-m,2S+2-n}\,,
\end{equation}
where $E_{m,n}$ are matrices of size $2S{+}1$ such that
$\bigl(E_{m,n}\bigr)_{kl}=\delta_{mk}\delta_{nl}$.

\subsection{Minimal polynomial ${\cal P}^{S,0}_{a_1,a_2}$ }\label{AMP}

For $\SH = a_1 \SP^{S,0}_{12} + a_2 \SP^{S,0}_{23}$ we have
\begin{align}\nonumber
 \SH^2 = a_1^2 \SP^{S,0}_{12} + a_2^2 \SP^{S,0}_{23} +
 a_1 a_2 (\SP^{S,0}_{12} \SP^{S,0}_{23} +
    \SP^{S,0}_{23} \SP^{S,0}_{12}) \,.
\end{align}
Multiplying this expression by $\SH$ and using \rf{TLa} we find
\begin{align}\nonumber
 \SH^3 = a_1^3 \SP^{S,0}_{12} + a_2^3 \SP^{S,0}_{23} +
 a_1 a_2 (a_1 \,{+}\, a_2) (\SP^{S,0}_{12} \SP^{S,0}_{23} +
    \SP^{S,0}_{23} \SP^{S,0}_{12}) + \mus \, a_1 a_2 \SH \,.
\end{align}
Whence
$ \SH^3 - (a_1 \,{+}\, a_2)\, \SH^2 = (\mus -1) \, a_1 a_2 \SH$.
Thus, the minimal polynomial for $\SH$ is~\rf{mP03}.

\subsection{Coefficients $d_k^{S,s}$ for minimal polynomials
${\cal P}^{S,s}_{a_1,a_2}$ }\label{CMP}

Let us denote $[t] \,{\equiv}\, \fr{\sin\gamma t}{\sin\gamma}$
and $\{t\} \,{\equiv}\, 2 \cos \gamma t$.
The coefficients $d_k^{S,s}$ in \rf{mP11} are given by
\begin{align}
\nonumber
 {}& S{=}1,\ s{=}1 \ : && d_1^{1,1} = \frac{1}{\{2\}^2} \,, \quad
  d_2^{1,1} = \Bigl(\frac{\{3\} }{\{1\} \{2\}} \Bigr)^2 ;\\
\nonumber
 {}& S{=}1,\ s{=}2 \ : &&
 d_1^{1,2} = \frac{1}{\{2\}^2} \,, \quad
 d_2^{1,2} = \Bigl(\frac{1 }{\{2\} [3]} \Bigr)^2 , \quad
 d_3^{1,2} = 1 ; \\
 \nonumber
 {}& S{=}\fr{3}{2},\ s{=}1 \ : &&
 d_1^{\frac{3}{2},1} = \Bigl(\frac{[3] }{\{2\} [5]} \Bigr)^2 , \quad
 d_2^{\frac{3}{2},1} = \frac{1}{\{2\}^2} \,, \quad
 d_3^{\frac{3}{2},1} = \Bigl(\frac{[2][6] \,{-}\, 1 }{[4] [5]} \Bigr)^2 ; \\
 \nonumber
 {}& S{=}\fr{3}{2},\ s{=}2 \ : &&
 d_1^{\frac{3}{2},2} = \frac{1}{\{3\}^2}  \,, \quad
 d_2^{\frac{3}{2},2} =  \frac{1}{\{2\}^2} \,, \quad
 d_3^{\frac{3}{2},2} =  \Bigl(\frac{\{5\} }{\{2\} \{3\}} \Bigr)^2 , \quad
 d_4^{\frac{3}{2},2} =  \Bigl(\frac{[5] \,{-}\, 2}{\{2\} \{3\}} \Bigr)^2 ; \\
 \nonumber
 {}& S{=}\fr{3}{2},\ s{=}3 \ : &&
 d_1^{\frac{3}{2},3} =  \frac{1}{\{3\}^2} \,, \quad
 d_2^{\frac{3}{2},3} =  \Bigl(\frac{ \{1\} }{\{3\} [5]} \Bigr)^2 , \quad
 d_3^{\frac{3}{2},3} =  \Bigl(\frac{ 1 }{\{2\}\{3\} [5]} \Bigr)^2 , \quad
 d_4^{\frac{3}{2},3} = 1 .
\end{align}
The minimal positive solutions $\gamma_{S,s}$ of the equation
$d_1^{S,s} \,{=}\, 1$ are the following:
\beq{ga1}
 \gamma_{1,1} = \gamma_{1,2} = \frac{\pi}{6} \,,\qquad
 \gamma_{\frac{3}{2},1} = \frac{\pi}{7} \,,\qquad
 \gamma_{\frac{3}{2},2} = \gamma_{\frac{3}{2},3} = \frac{\pi}{9} \,.
\end{equation}
Let us mention in passing an interesting pattern in the
minimal positive solutions of the equation
$d_k^{S,s} \,{=}\, 1$ for $s \,{=}\, 2S$:
we have $\gamma^{\klein \{1\}}_{1,2} \,{=}\, \fr{\pi}{6}$,
 $\gamma^{\klein \{2\}}_{1,2} \,{=}\, \fr{\pi}{5}$, and
$\gamma^{\klein \{1\}}_{\frac 32,3} \,{=}\, \fr{\pi}{9}$,
$\gamma^{\klein \{2\}}_{\frac 32,3} \,{=}\, \fr{\pi}{8}$,
$\gamma^{\klein \{3\}}_{\frac 32,3} \,{=}\, \fr{\pi}{7}$.

\subsection{Universal R--matrix }\label{AUR}
Drinfeld has shown \cite{D1} that relations \rf{uniRa} and
\rf{uniRb} are satisfied for $R^+$ and
$R^- \,{\equiv}\, \PP (R^+)^{-1} \PP$, where $R^+$ is given by
\beq{Ru}
  R^+ = q^{H \ot H} \, \sum_{n =0}^\infty
 \frac{ q^{\frac{1}{2}(n^2-n)} }{\prod_{k=1}^n [k]_q}
 \bigl( (q\,{-}\,q^{-1}) F \ot E \bigr)^n  \,
 q^{H \ot H} \,.
\end{equation}
Here $H$ is related to $K$ via $K\,{=}\,q^{H}$.
Relations \rf{uniRa}--\rf{uniRb} imply
the Yang--Baxter equation,
\beq{YB}
   R^\pm_{12} \, R^\pm_{13} \, R^\pm_{23} =
   R^\pm_{23} \, R^\pm_{13} \, R^\pm_{12} \,.
\end{equation}
Note that $ R^+\!\!\!\bigm|_{q\to q^{-1}}= \bigl(R^+\bigr)^{-1}$.
Therefore, for $|q| \,{=}\,1$ we have
\beq{R*}
   \bigl( R^+ \bigr)^* =  R^- \,.
\end{equation}

\subsection{Proof of Proposition~\ref{PEN}}\label{AP2}

Let us introduce an operation
$\Delta^\pm \,{\equiv}\, R^\pm \Delta$ and
define its action on $X \,{\in}\, U_q(sl_2)^{\ot N}$ by the
following formula:
$\Delta^\pm_{N,n}(X) \,{\equiv}\,
 R^\pm_{n,n+1} \Delta_{N,n}(X)$
(recall that $\Delta_{N,n}$ was defined after Eq.~\rf{HUsym}).

\begin{lem}\label{LemD}
a) $\Delta^\pm$ is coassociative, i.e.
\beq{RDel}
 \Delta^\pm_{2,1} \,{\circ}\, \Delta^\pm =
 \Delta^\pm_{2,2} \,{\circ}\, \Delta^\pm \,.
\end{equation}
Therefore, a positive integer power of $\Delta^\pm$ can
be defined in the same way as it is done for ${\Delta}$, i.e.
\beq{DelN}
 \bigl(\Delta^\pm\bigr)^{(N)} \,{=}\,
 \Delta^\pm_{N,n} \circ \bigl(\Delta^\pm\bigr)^{(N-1)} \,.
\end{equation}
The operations $\Delta^+$ and $\Delta^-$ are
conjugate to each other in the following sense:
\beq{Del*}
 \bigl(\Delta^+(X)\bigr)^* \,{=}\, \Delta^-(X^*) \,,
\end{equation}
for any $X \,{\in}\, U_q(sl_2)$.\\
b) The symmetrizing operators \rf{etaN} can be equivalently
represented as follows
\beq{etaNdel}
 \eta^\pm_{\klein N+1} =
  \pis^{\ot \klein N+1} \Bigl(
 \Delta^\pm_{N,n} \, \bigl(\tilde{\eta}^\pm_{\klein N} \bigr)
 \Bigr) = \pis^{\ot \klein (N+1)}
 \Bigl( \bigl(\Delta^\pm\bigr)^{(N)} \, (1) \Bigr) \,,
\end{equation}
where $\tilde{\eta}^\pm_{\klein N}$ are given by \rf{etaN}
with $R^\pm_{nm}$ instead of~$\SR^\pm_{nm}$, and
$\tilde{\eta}^\pm_1 \equiv 1$.

In \rf{DelN} and \rf{etaNdel}, $n$ can be taken
any from $1$ to $N$.
\end{lem}

\noindent
{\em Proof. a)}
The coassociativity of~$\Delta^\pm$ follows from the
coassociativity of $\Delta$ along with
the Yang--Baxter equation:
\begin{align*}
{}& \Delta^\pm_{2,1} \,{\circ}\, \Delta^\pm (X)=
 \Delta^\pm_{2,1} \bigl( R^\pm \Delta(X) \bigr) \={uniRb}
 R^\pm_{12} \, R^\pm_{13} \, R^\pm_{23} \, \Delta_{2,1}(X) \\
{}& \={YB}
 R^\pm_{23} \, R^\pm_{13} \, R^\pm_{12} \, \Delta_{2,2}(X) \={uniRa}
 \Delta^\pm_{2,2} \bigl( R^\pm \Delta(X) \bigr) =
 \Delta^\pm_{2,2} \,{\circ}\, \Delta^\pm (X) \,.
\end{align*}
The property \rf{Del*} is easily checked:
\begin{equation*}
  \bigl(\Delta^+(X)\bigr)^* = \bigl(R^+ \Delta(X)\bigr)^*
 \={R*} \Delta'(X^*) \, R^- \={uniRa} R^- \, \Delta(X^*)
 = \Delta^-(X^*) \,.
\end{equation*}

\noindent  {\em b)}
First, we will prove the first equality in \rf{etaNdel}
by an induction in the case of $n \,{=}\,N \,{-}\,1$.
The base of the induction, for $N\,{=}\,2$,
holds by the definition of $\Delta^\pm$ and the
relation $\Delta(1) \,{=}\, 1 \,{\ot}\, 1$.
The inductive step (which can be regarded as an
extension of the lattice by an additional node) is checked
as follows
\begin{align*}
{} \eta^\pm_{\klein N+1} & \={etaN}  \
 \stackrel{\leftarrow}{\SR}{\!}^\pm_{N+1}\!
 \stackrel{\leftarrow}{\SR}{\!}^\pm_{N} \, \eta^\pm_{\klein N-1} =
 \SR^\pm_{N,N+1} \SR^\pm_{N-1,N+1} \ldots  \SR^\pm_{1,N+1}
 \SR^\pm_{N-1,N} \ldots  \SR^\pm_{1N} \, \eta^\pm_{\klein N-1} \\
{}& = \SR^\pm_{N,N+1} (\SR^\pm_{N-1,N+1}\SR^\pm_{N-1,N} \,{\ldots}\,
  \SR^\pm_{n,N+1}\SR^\pm_{n,N}  \,{\ldots}\,
 \SR^\pm_{1,N+1}\SR^\pm_{1,N}) \,  \eta^\pm_{\klein N-1} \\
{}& \={uniRb} \pis^{\ot \klein (N+1)}
 \bigl( R^\pm_{N,N+1}
 \Delta_{N,N} (R^\pm_{N-1,N} \,{\ldots}\, R^\pm_{1N}) \,
 \tilde{\eta}^\pm_{\klein N-1} \bigr) \\
{}& = \pis^{\ot \klein (N+1)}
 \bigl(R^\pm_{N,N+1} \Delta_{N,N}
 (\stackrel{\leftarrow}{R}{\!}^\pm_{N} \,
 \tilde{\eta}^\pm_{\klein N-1}) \bigr) \={etaN}
 \pis^{\ot \klein (N+1)}
 \bigl( \Delta_{N,N}^\pm (\tilde{\eta}^\pm_{\klein N}) \bigr).
\end{align*}
Whence $\eta^\pm_{\klein N+1} \,{=}\,
 \pis^{\ot \klein (N+1)} \bigl(
 \Delta_{N,N}^\pm \circ \Delta_{N-1,N-1}^\pm
 \circ \dots \circ \Delta_{1,1}^\pm
 (\tilde{\eta}^\pm_{\klein 1}) \bigr)
 \={DelN} \pis^{\ot \klein (N+1)}
 \bigl( (\Delta^\pm)^{(N)}  (1) \bigr)$.
That is, we have proved the equality of $\eta^\pm_{\klein N+1}$
to the last expression in~\rf{etaNdel}. The latter in turn
is equal to the middle expression in \rf{etaNdel}, because
$n$ in the definition \rf{DelN} can be any from $1$ to~$N$.
This completes the proof of the Lemma~1.

{\em Proof} of  Proposition~\ref{PEN}.

\noindent
We commence by proving the part {\em b)}.
Choosing $n \,{=}\, 1$ in \rf{etaNdel}, we can
write $\eta^\pm_{\klein N+1}$  as follows:
$\eta^\pm_{\klein N+1} \,{=}\,
 \pis^{\ot \klein N+1}
 \bigl( (\Delta^\pm)^{(N)}({1}) \bigr)
 = \pis^{\ot \klein N+1}
 \bigl( \Delta_{N,1}^\pm \circ \Delta_{N-1,1}^\pm \circ
 \dots \circ \Delta_{1,1}^\pm (1) \bigr)$. Then
expressions \rf{etaNop} can be obtained by an induction
analogous to that was performed in the proof of Lemma~1
but this time one should use the first relation in~\rf{uniRb}.

Relation \rf{eta*} in the part {\em c)} of Proposition~\ref{PEN}
is an immediate consequence of applying relation \rf{Del*}
to formula~\rf{etaNdel}.

To prove the part {\em a)} of Proposition~\ref{PEN},
we show first that $\eta^\pm_{\klein N}$ are symmetrizing
operators for the tensor Casimir element:
\begin{align*}
{}& \eta^\pm_{\klein N} \, \SC_{n,n+1} \={etaNdel}
 \pis^{\ot \klein N}
 \bigl( R^\pm_{n,n+1} \Delta_{N-1,n}
 (\tilde{\eta}^\pm_{\klein N-1} \, C_n) \bigr)
 = \pis^{\ot\klein N}
 \bigl( R^\pm_{n,n+1} \Delta_{N-1,n}
 (C_n \, \tilde{\eta}^\pm_{\klein N-1} ) \bigr) \\
{}& \={etaNdel}
 \SR^\pm_{n,n+1} \, \SC_{n,n+1} \, \bigl(\SR^\pm_{n,n+1})^{-1} \,
  \eta^\pm_{\klein N}
  \={uniRa} \SC_{n+1,n} \, \eta^\pm_{\klein N} \,.
\end{align*}
Therefore $\eta^\pm_{\klein N}$ are symmetrizing operators
also for an arbitrary polynomial in $\SC_{n,n+1}$ with
real coefficients. Whence, taking formula \rf{PjC} into account,
we conclude that relation \rf{etaP} holds.
Thus, Proposition~\ref{PEN} is proven.

\subsection{Proof of Lemma~\ref{DE}}\label{AP4}
The bialgebra defined by relations (16)--(17)
turns into a Hopf algebra if the antipode ${\cal S}$ (an
antihomomorphism) is defined as follows:
${\cal S}(E) \,{=}\, -q^{-1}E$, ${\cal S}(F) \,{=}\, -q F$,
${\cal S}(K) \,{=}\, K^{-1}$.

The R--matrix \rf{Ru} has the following form:
$R^+ \,{=}\, \sum_a r^{\klein(1)}_a \,{\ot}\, r^{\klein(2)}_a$.
Consider the element
$\chi \,{=}\, K^2 \, \bigl(\sum_a {\cal S}(r^{\klein(2)}_a)
 r^{\klein(1)}_a \bigr)$. From the results of \cite{D2}, it follows
that $\chi$ is a central element, which acquires the value
$q^{-2S(S+1)}$ on an irreducible module $V^S$, and that
$\chi$ satisfies the following relation:
\beq{chidel}
 \chi_1 \, \chi_2 \, \Delta(\chi^{-1}) =
 \bigl(R^- \bigr)^{-1} \, R^+ \,.
\end{equation}

Let us prove that
\beq{etachi}
 \chi_1 \,{\ldots}\, \chi_{\klein N} \,
    \Delta^{\klein(N-1)}(\chi^{-1}) =
 \bigl(\tilde{\eta}^-_{\klein N} \bigr)^{-1} \,
    \tilde{\eta}^+_{\klein N} \,.
\end{equation}
For $N \,{=}\,2$, this relation coincides with \rf{chidel}.
For $N \,{\geq}\,3$, it is verified by induction:
\begin{align*}
 {}& \chi_1 \,{\ldots}\, \chi_{\klein N+1} \,
    \Delta^{\klein(N)}(\chi^{-1}) \={chidel}
  \bigl(R_{12}^- \bigr)^{-1} R_{12}^+ \,
 \Delta_{\klein N,1} \bigl( \chi_1 \,{\ldots}\, \chi_{\klein N} \,
    \Delta^{\klein(N-1)}(\chi^{-1}) \bigr) \\
 {}& \={etachi} \bigl(R_{12}^- \bigr)^{-1} R_{12}^+ \,
\Delta_{\klein N,1} \bigl( (\tilde{\eta}^-_{\klein N})^{-1} \bigr) \,
 \Delta_{\klein N,1} \bigl( \tilde{\eta}^+_{\klein N} \bigr)
{}= \bigl( \Delta^-_{\klein N,1}(\tilde{\eta}^-_{\klein N}) \bigr)^{-1}
 \, \Delta^+_{\klein N,1}(\tilde{\eta}^+_{\klein N})
{} \={etaNdel} \bigl(\tilde{\eta}^-_{\klein N+1} \bigr)^{-1} \,
    \tilde{\eta}^+_{\klein N+1} .
\end{align*}

If $q$ is not a root of unity, the center of the algebra $U_q(sl_2)$
is generated by the Casimir element~\rf{C}.
Therefore, there exists a function $\varphi_q$ such that
$\chi \,{=}\, \varphi_q(C)$. Consequently, the operator
$\Delta^{\klein(N-1)}(\chi) \,{=}\,
\varphi(\Delta^{\klein(N-1)}(C))$
acts in each irreducible submodule
$V^s \,{\subset}\, (V^S)^{\klein \ot N}$ as multiplication by
$q^{-2s(s+1)}$. This, along with formula \rf{etachi},
implies that
\beq{etaPs}
 \bigl(\eta^-_{\klein N} \bigr)^{-1} \eta^+_{\klein N} =
 \sum_{s= s_\0}^{NS} q^{2s(s+1)-2NS(S+1)} \, {\cal P}_s \,,
\end{equation}
where ${\cal P}_s$ denotes the projector of rank
$\nu_s(2s\,{+}\,1)$ onto the reducible invariant subspace
$\oplus^{\nu_s} V^s\,{\subset}\, (V^S)^{\klein \ot N}$.

Using \rf{etaPs}, we derive formula \rf{specRN}:
\begin{align*}
{}&   \det\bigl( e^{i\alpha} \eta^+_{\klein N} +
    e^{-i\alpha} \eta^-_{\klein N} \bigr) =
 \det(\eta^-_{\klein N}) \,
 \det\bigl( e^{i\alpha} (\eta^-_{\klein N})^{-1}
    \eta^+_{\klein N} + e^{-i\alpha} \, \mathsf{1} \bigr) \\
{}& \={etaPs} \det\Bigl( \sum_{s= s_\0}^{NS} (e^{i\alpha}
     q^{2s(s+1)-2NS(S+1)} + e^{-i\alpha})\, {\cal P}_s \Bigr) \\
{}& \ = \rho_{\klein N,S} \,
 \prod_{s=s_\0}^{SN} \Bigl(e^{i\alpha} q^{s(s+1)-NS(S+1)} +
 e^{-i\alpha} q^{NS(S+1) -s(s+1)} \Bigr)^{\nu_s(2s+1) } \,,
\end{align*}
where $\rho_{\klein N,S} \,{\equiv}\, \prod_{s=s_\0}^{SN}
    q^{\nu_s(2s+1)(s(s+1)-NS(S+1))} \,{=}\,1$,
which follows from \rf{etaPs} and the relation
$\det \eta^\pm_{\klein N} \,{=}\,1$
(note that $\det\SR^\pm \,{=}\, 1$).

\par\vspace*{3mm}\noindent
{\small
{\bf Acknowledgements.}
 This work was started during author's visit (supported
 by the Swiss National Science Foundation under grant 200020--121675)
 to Mathematics Department, University of Geneva, and
 completed during a visit (supported by the Alexander von
 Humboldt Foundation) to DESY, Hamburg.
 The work was also supported in part by the Russian Foundation
 for Fundamental Research (grants 07--02--92166, 08--01--00638,
 09--01--12150, 09--01--93108).

 The author thanks A.~Alekseev, A.~Fring, G.~von Gehlen,
 P.~Kulish, and V.~Tarasov for useful remarks.
}

%
\def\baselinestretch{1}

\end{document}